\documentclass[12pt]{iopart}
\begin{document}

\title[Superradiance from an ultrathin film of three-level
$V$-type atoms]{Superradiance from an ultrathin film of
three-level $V$-type atoms: Interplay between splitting, quantum
coherence and local-field effects}

\author{ V. A. Malyshev\dag\footnote[3]{On leave from "S.I.
Vavilov State Optical Institute", Birzhevaya Linia 12, 199034
Saint-Petersburg, Russia} F. Carre\~{n}o\ddag\ M. A.
Ant\'{o}n\ddag\ Oscar G. Calder\'on\ddag\ and F.
Dom\'{\i}nguez-Adame\dag}

\address{\dag\ GISC, Departamento de F\'{\i}sica de Materiales,
Universidad Complutense, 28040 Madrid, Spain}

\address{\ddag\ Escuela Universitaria de \'Optica,
Universidad Complutense de Madrid, C/ Arcos de Jal\'on s/n, 28037
Madrid, Spain}

\eads{\mailto{vicmal@valbuena.fis.ucm.es},
\mailto{fcarreno@ucmail.ucm.es}, \mailto{antonm@fis.ucm.es},
\mailto{oscargc@opt.ucm.es}, \mailto{adame@valbuena.fis.ucm.es}}

\begin{abstract}

We carry out a theoretical study of the collective spontaneous
emission (superradiance) from an ultrathin film comprised of
three-level atoms with $V$-configuration of the operating
transitions. As the thickness of the system is small compared to
the emission wavelength inside the film, the local-field
correction to the averaged Maxwell field is relevant. We show that
the interplay between the low-frequency quantum coherence within
the subspace of the upper doublet states and the local-field
correction may drastically affect the branching ratio of the
operating transitions. This effect may be used for controlling the
emission process by varying the doublet splitting and the amount
of low-frequency coherence.

\end{abstract}

\pacs{42.50.Fx, 42.50.Md}

\submitto{\JOB}

\maketitle

\section{Introduction}
\label{Intro}

After the pioneering work by Kocharovskaya and
Khanin~\cite{Kocharovskaya88,Khanin90}, Harris~\cite{Harris89},
and Scully, Zhu and Gavrielides~\cite{Scully89}, the effects of
the light-matter interactions in the presence of low-frequency
quantum coherence between sublevels in the ground or excited
states have received much attention. There appeared new phenomena
such as {\it amplification\/} and {\it lasing without
inversion\/}~(AWI and LWI), {\it electromagnetically induced
transparency\/}~(EIT), etc.\@ (see
references~\cite{Kocharovskaya92,Mompart00} for review as well as the
topical issue of Quantum Optics~\cite{QO94}). Optical bistability in
$V$-type atoms has been considered in reference \cite{Anton02}.

Recently, it was shown that in a close system of $\Lambda$-type
atoms, the low-frequency coherence may give rise to superradiance
without inversion
(SRWI)~\cite{Malyshev97a,Manassah98,Zaitsev99a,Zaitsev99b,Zaitsev01,Ryzhov01}.
In reference~\cite{Kozlov99}, the SRWI of an open (i.e.\@ in the
presence of a driven field) system of $V$-type atoms was
discussed. We might stress that the standard superradiance (SR)
effect, predicted by Dicke for a collection of two-level
atoms~\cite{Dicke54}, requires an initial inversion of level
populations (see the review by Gross and Haroche~\cite{Gross82}
and the book by Benedict {\it et al.}~\cite{Benedict96} for futher
details).

The goal of the present paper is to analyze theoretically the
features of the SR of a close system of $V$-type atoms. This
problem was already discussed in the eighties by Crubellier {\it
et al}~\cite{Crubellier80,Crubellier85} and by Molander and
Stroud~\cite{Molander82} within the context of subradiance
---collective spontaneous emission from  weakly radiating collective states---.
This  effect has been observed experimentally by Pavolini {\it et
al.}~\cite{Pavolini85} in a low-density gas of gallium atoms (the
density of gallium atoms in the interaction region was about
$10^{12}\,$atoms/cm$^3$). Here, we consider a dense system
comprised of $V$-type radiators and take into account the
local-field correction (LFC) to the averaged Maxwell field. It
should be stressed that this correction was neglected in previous
theoretical studies~\cite{Crubellier80,Crubellier85,Molander82},
and this approach  is then the novelty of the paper. As is well
known from the studies of high-density two-level
assemblies~\cite{Friedberg73,Hopf84,Ben-Aryeh86,Stroud88,Friedberg89,Crenshaw92a,Crenshaw92b,Scalora95,Crenshaw96,Manassah96,Malyshev97b,Jarque97,Afanas'ev98,Maimistov99},
the LFC affects significantly the resonant optical response of
such systems, especially the response of ultrathin
films~\cite{Benedict89,Samson90,Benedict91,Oraevsky94}. LFC in a
$V$-type medium in the non-cooperative regime has been analyzed in
reference~\cite{Calderon03}. In particular, the effects of the LFC
on the SR of a $\Lambda$-type dense system have been discussed in
references~\cite{Zaitsev99b,Zaitsev01,Ryzhov01,Malyshev00}. We
show that in the case of a dense $V$-type system, the interplay
between the LFC, doublet splitting and low-frequency coherence
dramatically influences the branching ratio of the operating
transitions. This efect can be used for controlling the SR by
varying the doublet splitting and the degree of coherence between
the doublet states.

The outline of the paper is as follows. In section~\ref{Model} we
present the model we will be dealing with. We derive the truncated
equations for the density matrix elements and electric field
taking into account the LFC to the Maxwell field, within the
rotating wave approximation (RWA). Section~\ref{Analytics} is
devoted to the particular case of degenerate doublet in the upper
state, allowing for an analytical solution of the problem. Results
of numerical simulations are described in section~\ref{Numerics},
where we provide as well interpretations of the peculiarities
found numerically. We conclude the paper in section~\ref{Summary}
with a brief summary of the results and how these can be related
to actual measurements to infer the main characteristics of the
LFC.

\section{Model and truncated equations}
\label{Model}

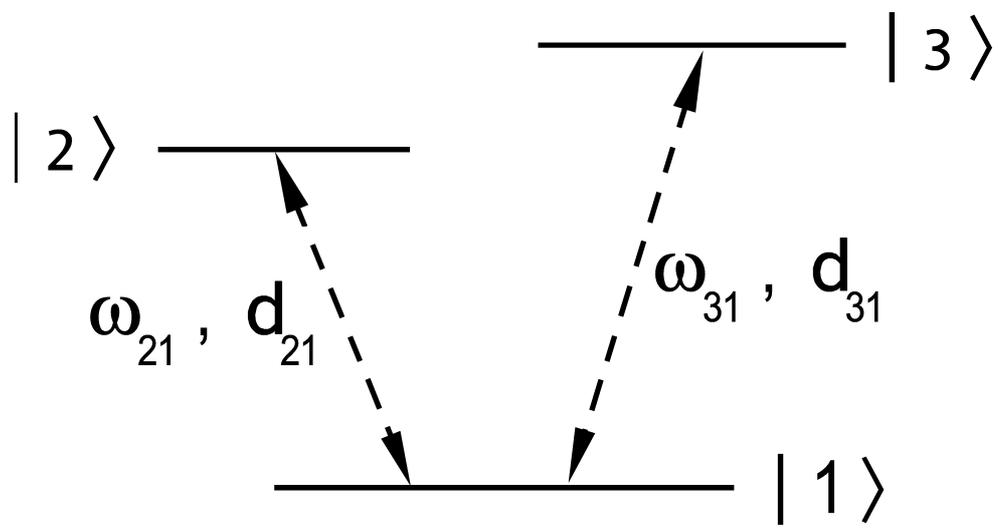
\begin{figure}[ht]
\caption{Scheme of the energy levels and transitions in a $V$-type
atom.} \label{fig1}
\end{figure}

We consider a system of three-level atoms with a doublet in the
upper state (the so-called $V$-configuration, see
figure~\ref{fig1}), forming an ultrathin film of thickness $L$
smaller than the emission wavelength inside the film. This
constraints is usually met in experiments and simplify the
mathematical description of the model by taking away the
propagation effect along the film normal. All the vectors, the SR
electric field and transition dipole moments of the operating
transitions, ${\bf d}_{21}$ and ${\bf d}_{31}$, are considered to
be parallel to each other as well as to the film plane (the
transition between the levels of the doublet is not included).
Under these assumptions, the vector nature of the above quantities
is not important, and the problem is reduced to its scalar form.
It is also assumed, without loss of generality, that the
transition dipole moments are real and positive: $d_{21} = d_{12}
> 0$ and $d_{31} = d_{13} > 0$.

We will consider the semiclassical description of the optical
dynamics of the system, treating the atom dynamics by means of the
density matrix $\rho_{\alpha\beta} \ (\alpha, \beta = 1,2,3)$,
i.e., quantum mechanically, while the field evolution is accounted
for the Maxwell equation, i.e., classically. The effect of quantum
coherence on the spontaneous emission of a single atom is not
taken into account since is much weaker than the collective
effect.

Under the limitations adopted above, the set of equations for the
joint system of the density matrix-Maxwell equations reads
\begin{eqnarray}
{\dot \rho}_{31} & = & - i \omega_{31} \rho_{31} -
            i \frac{d_{31} {\cal E}^\prime}{\hbar} (\rho_{33}
            - \rho_{11})
            - i \frac{d_{21} {\cal E}^\prime}{\hbar} \rho_{32} \ ,
\label{1rho31} \\
{\dot \rho}_{21} & = & - i \omega_{21} \rho_{21} -
            i \frac{d_{21} {\cal E}^\prime}{\hbar} (\rho_{22}
            - \rho_{11})
            - i \frac{d_{31} {\cal E}^\prime}{\hbar} \rho_{23} \ ,
\label{1rho21} \\
{\dot \rho}_{32} & =  & - i \omega_{32} \rho_{32} -
            i \frac{d_{21} {\cal E}^\prime}{\hbar} \rho_{31} +
            i \frac{d_{31} {\cal E}^\prime}{\hbar} \rho_{12} \ ,
\label{1rho32} \\
{\dot \rho}_{33} & =  & i \frac{d_{31} {\cal E}^\prime}{\hbar}
(\rho_{13} -
            \rho_{31}) \ ,
\label{1rho33} \\
{\dot \rho}_{22} & = & i \frac{d_{21} {\cal E}^\prime}{\hbar}
(\rho_{12} -
            \rho_{21}) \ ,
\label{1rho22} \\
{\dot \rho}_{11}  & = & - i \frac{d_{21} {\cal E}^\prime}{\hbar}
(\rho_{12} -
            \rho_{21})
                   - i \frac{d_{31} {\cal E}^\prime}{\hbar} (\rho_{13} -
            \rho_{31})\ .
\label{1rho11}
\end{eqnarray}

\vskip 0.5cm \noindent Here, the dot denotes time derivative and
${\cal E}^\prime$ stands for the acting field
\begin{equation}
{\cal E}^\prime = - \frac{2\pi L}{c}\dot{\cal P} + \frac{4\pi}{3}
{\cal P}\ , \label{Eprime}
\end{equation}
where $L$ and $c$ denote the film thickness and the speed of
light, respectively, and ${\cal P} = N_0 (d_{31} \rho_{31} +
d_{21} \rho_{21} + \mathrm{c.c.})$ is the electric polarization of
the unit volume, with $N_0$ being the atom number concentration.
The first term in equation~(\ref{Eprime}) represents the
Maxwellian emission field, while the second term is the LFC.

In order to further specify the model we are dealing with, we
first stress that it is applicable to the description of SR of a
thin dielectric crystalline film rather than a dense gas system.
In the latter case, the pressure broadening terms turn out to be
important. Having the same nature and order of magnitude as the
LFC~\cite{Friedberg89}, these terms have to be added to the
equations for the off-diagonal density matrix elements. In solid
crystalline media, the levels are usually broadened due to crystal
imperfections as well as coupling to phonons. Under specific
conditions, the width of the levels is smaller than the LFC (in
frequency units). The existence of Frenkel exciton states in
dielectric solids, which are due to the interatomic dipolar
coupling~\cite{Davydov71,Agranovich82} or, in other words, due to
the LFC, represents an unambiguous confirmation of this fact.
Because of that, we do not take into account either relaxation of
populations or dephasing of the electric polarization of a single
atom, assuming that the SR process is faster (the estimates of the
corresponding constants are presented in section~\ref{Summary}).

We seek a solution of equations~(\ref{1rho31})-(\ref{1rho11}) in
the form: ${\cal E}^\prime = E^\prime \exp(-i\omega_{c}t) +
\mathrm{c.c.}$,\ $\rho_{31}=R_{31}\exp(-i\omega_{c}t)$,\
$\rho_{21}= R_{21}\exp(-i\omega_{c}t)$,\ where $\omega_c =
(\omega_{31}+\omega_{21})/2$; $E^\prime$ and $R_{31}, R_{21}$ are
the complex slowly varying (in the scale $2\pi/\omega_c$)
amplitudes of the field and of the off-diagonal density matrix
elements, respectively.  Hereafter the latters will be referred to
as optical coherences. Within the RWA, the equations for the
amplitudes read
\begin{eqnarray}
{\dot R}_{31} & = & -i\frac{\omega_{32}}{2}R_{31} +
        \left( \frac{1}{\tau_{R}} - i\Delta_{L} \right)
        [\mu_{31}(\rho_{33} - \rho_{11})
 \nonumber \\
& &
        + \mu_{21}\rho_{32}]
        (\mu_{21}R_{21} + \mu_{31}R_{31}) \ ,
\label{2R31} \\
{\dot R}_{21}  & = & i\frac{\omega_{32}}{2}R_{21} +
        \left(\frac{1}{\tau_{R}} - i\Delta_{L}\right)
        [\mu_{21}(\rho_{22}-\rho_{11})
\nonumber \\
& &
        + \mu_{31}\rho_{23}]
        (\mu_{21}R_{21}+\mu_{31}R_{31}) \ ,
\label{2R21} \\
{\dot\rho}_{32} & = & -i\omega_{32}\rho_{32} - \left[ \left(
\frac{1}{\tau_R} +
        i\Delta_L\right) \mu_{21}R_{31} (\mu_{21}R_{21}^* + \mu_{31}R_{31}^*)
\right.
\nonumber \\
 & & \left.
       + \left( \frac{1}{\tau_R} - i\Delta_L\right) \mu_{31}R_{21}^*
       (\mu_{21}R_{21} + \mu_{31}R_{31}) \right] \ ,
\label{2rho32} \\
{\dot\rho}_{33} &  = &  \mu_{31} \left[ \left( -
\frac{1}{\tau_{R}}
       + i \Delta_L \right)(\mu_{21}R_{21} + \mu_{31}R_{31})R_{31}^*
       + \mathrm{c.c.} \right] \ ,
\label{2rho33} \\
{\dot\rho}_{22} & = & \mu_{21} \left[ \left( - \frac{1}{\tau_{R}}
       + i \Delta_L \right)(\mu_{21}R_{21} + \mu_{31}R_{31})R_{21}^*
       + \mathrm{c.c.} \right] \ ,
\label{2rho22} \\
{\dot\rho}_{11} & = & \frac{2}{\tau_{R}}|\mu_{21}R_{21} +
        \mu_{31}R_{31}|^2 \ .
\label{2rho11}
\end{eqnarray}
Here we have defined $\mu_{31} = d_{31}/d$  and $\mu_{21} =
d_{21}/d$,  where $d = \sqrt{(d_{31}^2 + d_{21}^2)/2}$; \
$\Delta_{L} = 4\pi d^2N_0/3\hbar$; \ $\tau_{R}^{-1}=2\pi k_{c}L
d^{2}N_{0}/\hbar, \ k_c = \omega_c/c$.  When deriving
equations~(\ref{2R31})-(\ref{2rho11}), we exploited the fact that
the equation for the slowly-varying field amplitude $E^\prime$ can
be cast in the form
\begin{equation}
\frac{dE^\prime}{\hbar} =  \left(\frac{i}{\tau_R} +\Delta_L\right)
        (\mu_{21}R_{21} + \mu_{31}R_{31}) \ ,
\label{1Eprime}
\end{equation}
and we introduced this expression directly in the density matrix
equation. The quantities $\tau_R^{-1}$ and $\Delta_L$ represent
the magnitudes of the SR field and of the LFC (in frequency
units), respectively~\cite{Benedict91}. Recall that $\Delta_L >
\tau_R^{-1}$ since the relationship  $k_c L < 1$ holds for an
ultrathin film.

It is to be noted that equations~(\ref{2R31})-(\ref{2rho11}) have
the following integrals of motion:
\begin{eqnarray}
\rho_{11} + \rho_{22} + \rho_{33} = 1 \ , \label{Norm} \\
\rho_{11}^2 + \rho_{22}^2 +
\rho_{33}^2+2(|\rho_{32}|^2+|R_{31}|^2+|R_{21}|^2) =
\mathrm{const} \ , \label{???}
\end{eqnarray}
where the first equation establishes the normalization condition
for the total level population, while the second one cannot be
interpreted in a simple way.

To complete the mathematical formalism we should specify the
initial conditions for equations~(\ref{2R31})-(\ref{2rho11}). We
assume that the doublet states are initially populated, i.e.,
there exist nonzero $\rho_{33}(0)$ and $\rho_{22}(0)$. We also
allow an initial low-frequency coherence $\rho_{32}(0)$. In order
to trigger the emission process, we set a fixed (not-fluctuating)
value for the initial electric polarization in the operating
channels, $R_{31}(0) = R_{21}(0) = R_0$. This corresponds to
triggering the SR by an ultrashort external pulse of a small area,
with a duration $T_p < \mathrm{min} \{2\pi/\omega_{32},
\tau_R^{-1}\}$~\cite{Benedict96,Carlson80,Malikov84}.

Equations~(\ref{2R31})-(\ref{2rho11}) are written within the
original basis of states $|1 \rangle$, $|2 \rangle$ and $|3
\rangle$. From physical reasons, especially  in the case of a
degenerated doublet (see below), another set of states turns  out
to be very useful: $|1 \rangle$, $|+ \rangle =
(1/\sqrt{2})(\mu_{21}|2 \rangle + \mu_{31}|3\rangle)$ and $|-
\rangle = (1/\sqrt{2}) (\mu_{21}|3\rangle - \mu_{31}|2\rangle )$.
The convenience of this set is clear from the fact that only  the
superposition $|+ \rangle$ is coupled to the ground state $|1
\rangle$ (it will be referred to as bright state hereafter), while
the remainder  one is decoupled (dark state). The dipole moments
of the transitions $|1 \rangle \rightarrow |+ \rangle$ and  $|1
\rangle \rightarrow |- \rangle$ are $\langle 1|{\hat d}|+\rangle =
\sqrt{2}d$  and $\langle 1|{\hat d}|-\rangle = 0$. In this
regards, the dark channel does not contribute to the SR.

Within the new basis, $|1\rangle , \ |+\rangle , \ |-\rangle$, the
density matrix  elements can be expressed as follows
\begin{eqnarray}
R_{+1} & = & \frac{1}{\sqrt 2}(\mu_{21}R_{21} + \mu_{31}R_{31}) \
, \label{3R+1} \\
\rho_{++} & = & \frac{1}{2}(\mu_{21}^{2}\rho_{22} +
\mu_{31}^{2}\rho_{33}  +
        2\mu_{21}\mu_{31}\Re \rho_{32})\ ,
\label{3rho++} \\
R_{-1} & = & \frac{1}{\sqrt 2}(\mu_{21}R_{31} - \mu_{31}R_{21}) \
, \label{3R-1} \\
\rho_{--} & = & \frac{1}{2}(\mu_{21}^{2}\rho_{33} +
\mu_{31}^{2}\rho_{22} -
        2\mu_{21}\mu_{31} \Re \rho_{32}) \ ,
\label{3rho--} \\
\rho_{+-} & = & \frac{1}{2}[\mu_{21}\mu_{31}(\rho_{33}-\rho_{22})
+       \mu_{21}^{2}\rho_{23} - \mu_{31}^{2}\rho_{32}] \ ,
\label{3rho+-}
\end{eqnarray}
where now $\rho_{++}$ and $\rho_{--}$ stand for populations of the
bright and dark states, respectively; $\rho_{+-}$ represents the
low-frequency coherence, while $R_{+1}$ and $R_{-1}$ describe the
coherence of the bright and dark channels,  respectively. We
stress that $R_{+1}$ determines the field, as seen from
equation~(\ref{Eprime}). The equations for these matrix elements
read
\begin{eqnarray}
{\dot R}_{+1}  & = &
        -i\frac{\omega_{32}}{4}\Bigl[(\mu_{31}^2 - \mu_{21}^{2})R_{+1}
        + 2\mu_{21}\mu_{31}R_{-1}\Bigr]
\nonumber \\
& &
        + 2\left(\frac{1}{\tau_R} - i\Delta_{L}\right)(\rho_{++}
        - \rho_{11})R_{+1} \ ,
\label{4R+1} \\
{\dot\rho}_{++} & = &
        i\frac{\omega_{32}}{2}\mu_{21}\mu_{31}\Bigl(\rho_{+-}-\rho_{-+}\Bigr)
         -\frac{4}{\tau_{R}}|R_{+1}|^{2} \ ,
\label{4rho++} \\
{\dot\rho}_{11} & = & \frac{4}{\tau_R}|R_{+1}|^{2} \ ,
\label{4rho11} \\
{\dot R}_{-1} & = &
        - i\frac{\omega_{32}}{4}\Bigl[(\mu_{21}^2 - \mu_{31}^{2})R_{-1}
        + 2\mu_{21}\mu_{31}R_{+1}\Bigr]
        + 2\left(\frac{1}{\tau_R} - i\Delta_{L}\right)R_{+1}\rho_{-+} \ ,
\label{4R-1} \\
{\dot\rho}_{+-} & = &
        i\frac{\omega_{32}}{2}\Bigl[(\mu_{21}^{2}-\mu_{31}^{2})\rho_{+-}
        + \mu_{21}\mu_{31}(\rho_{++} - \rho_{--})\Bigr]
\nonumber \\
& &
        +     2\left(- \frac{1}{\tau_R} + i\Delta_{L}\right)R_{+1}R_{-1}^* \ ,
\label{4rho+-} \\
{\dot\rho}_{--} & = & i\frac{\omega_{32}}{2}\mu_{21}\mu_{31}
        \left(\rho_{-+} - \rho_{+-}\right)\ .
\label{4rho--}
\end{eqnarray}
As can be seen from equations~(\ref{4R+1})-(\ref{4rho--}), the
bright channel ($|+\rangle \rightarrow |1\rangle$) is coupled to
the dark one ($|- \rangle \rightarrow |1\rangle$) through
$\omega_{32}$--terms, and thus, at $\omega_{32} = 0$, the former
turns out to be independent of the latter (see
section~\ref{Analytics} for more details). At the same time, the
behavior of the dark channel is driven by the the bright one even
in the presence of degeneracy of the doublet states: the field
terms  proportional to $\tau_R^{-1}$ and $\Delta_L$ play this
role.

\section{Degenerated doublet ($\omega_{32} = 0$)}
\label{Analytics}

We first analyze the degenerated case. Then
equations~(\ref{4R+1})-(\ref{4rho--}), describing the bright
channel and which we are interesting in, reduce to
\begin{eqnarray}
\frac{d}{dt} |R_{+1}| & = &
         \frac{4}{\tau_R}\; Z\; |R_{+1}| \ ,
\label{5R+1} \\
{\dot Z} & = &
         -\frac{4}{\tau_{R}}\; |R_{+1}|^{2} \ ,
\label{5Z} \\
{\dot \phi} & = &
         -4\Delta_{L}\,Z\ ,
\label{5phi}
\end{eqnarray}
where the new variables read $Z \equiv (\rho_{++} - \rho_{11})/2$,
and $\phi$ is  the phase of $R_{+1}$. We stress that
equations~(\ref{5R+1})-(\ref{5phi}) are similar  to the SR
equations of an ultrathin film of two-level
atoms~\cite{Benedict96,Stroud72,Zaitsev83,Avetisyan85}, but
replacing $\tau_R$ by $\tau_R/2$ and $\Delta_L$ by $2\Delta_L$,
which is reasonable due to the presence of two emission channels.

Several qualitative conclusions about the system behavior can be
drawn from the direct analysis of
equations~(\ref{5R+1})-(\ref{5phi}). First of all, the derivative
$d|R_{+1}|/dt$ is positive and thus $|R_{+1}|$ will give rise to
the SR if $Z(0)  > 0$ or, in other words, if there is an initial
population inversion between the bright and ground states:
\begin{equation}
\rho_{++}(0) =
        \frac{1}{2}\Big[ \mu_{21}^{2}\rho_{22}(0) + \mu_{31}^{2}\rho_{33}(0)  +
        2\mu_{21}\mu_{31}\Re \rho_{32}(0)\Big] \; > \; \rho_{11}(0)\ .
\label{Z>0}
\end{equation}

From here it can be shown that, in order to meet this inequality,
the total initial population of the doublet, $\rho_{22}(0) +
\rho_{33}(0)$, must by larger than the population in the ground
state, $\rho_{11}(0)$. In other words, in contrast to the case of
a close $\Lambda$-system where the SRWI can be
observed~\cite{Malyshev97a}, the SR in a close $V$-type system
requires the population  inversion between actual levels. For the
sake of simplicity, let us set $\rho_{22}(0) = \rho_{33}(0) =
\rho_{32}(0) \equiv A = (1/2) [1 - \rho_{11}(0)]$. It corresponds
to the excitation of the bright state $|+ \rangle = (1/\sqrt{2})
(\mu_{21}|2 \rangle + \mu_{31}|3 \rangle)$ with amplitude $A$.
Then, the inequality~(\ref{Z>0}) takes the form $A(1 +
\mu_{21}\mu_{31}) > \rho_{11}(0)$. Bearing in mind that
$\mu_{21}\mu_{31} < 1$, we get $2A > \rho_{11}(0)$, i.e., the
total population of the doublet must to be indeed larger than that
in the ground state.

With the substitutions $Z = B\cos\Theta$, and $|R_{+1}| = B
\sin\Theta$, where $B \approx Z(0)$,
equations~(\ref{5R+1})-(\ref{5phi}) can be solved analytically,
and the solution is
\begin{eqnarray}
    Z & = & - Z(0) \tanh\left(\frac{t-t_D}{\tau_R^\prime}\right) \ ,
\label{6Z} \\
    |R_{+1}| & = & Z(0) {\rm sech} \left(\frac{t-t_D}{\tau_R^\prime} \right) \ ,
\label{6R+1} \\
    \phi & = & -4 \Delta_L \int_0^t Z(\tau) d\tau =
    4Z(0)\Delta_L\tau_R^\prime \left[ \ln \cosh \left(\frac{t-t_{D}}
    {\tau_R^\prime}\right)
\right. \nonumber \\
 & & \left.
    -\ln\cosh\left(\frac{t_D}{\tau_R^\prime}\right)
    \right] \ ,
\label{6phi} \\
    t_D & = & \tau_R^\prime \ln \left[\frac{2Z(0)}{|R_{+1}(0)|} \right] \ .
\label{6tD}
\end{eqnarray}
As is seen, the SR pulse is characterized by a delay time $t_D$
and a duration $\tau_R^\prime = \tau_R/4Z(0)$. The unique effect
of the LFC on the SR from a degenerated $V$-system is the SR phase
modulation which changes the SR frequency
\begin{equation}
    \Omega(t) = {\dot\phi} = 4Z(0)\; \Delta_L \tanh\left(\frac{t-t_D}
    {\tau_R^\prime}\right) \ ,
\label{Omega}
\end{equation}
from $-4Z(0)\Delta_L$ to $4Z(0)\Delta_L$. For this reason, it is
similar to what is well known for two-level dense
systems~\cite{Benedict96,Stroud72,Zaitsev83,Avetisyan85}, namely
within the mean-field approximation we are in fact dealing with,
the LFC does not affect the SR kinetics, but determines the width
of the SR spectrum.

\section{Nondegenerated doublet}
\label{Numerics}

We show below that the scenario of the SR from a nondegenerated
$V$-system changes dramatically in the presence of the LFC. It is
to be noticed that, concerning the SR from a $\Lambda$-system,
this fact has been already mentioned in
references.~\cite{Zaitsev99a,Zaitsev99b,Malyshev00,Zaitsev01,Ryzhov01}.
In order to investigate systematically the peculiarities of the SR
in the case of a $V$-system,  we perform the numerical solution of
equations~(\ref{2R31})-(\ref{2rho11}). In all our calculations,
the dipole moments of the operating transitions $|3 \rangle
\rightarrow | 1 \rangle $ and $ | 2 \rangle \rightarrow | 1
\rangle $ are equal to each other, thus implying that $\mu_{21} =
\mu_{31} = 1$. The initial values of amplitudes of the
high-frequency coherences are set to $R_{31}(0) = R_{32}(0) =
10^{-8}$. Time is expressed in units of $\tau_R$. The other
initial magnitudes, such that the level populations $\rho_{11}(0),
\> \rho_{22}(0)$ and \, $\rho_{33}(0)$, the low-frequency
coherence $\rho_{32}(0)$, the doublet splitting $\omega_{32}$ and
the LFC $\Delta_L$, will be regarded as variable parameters.

\subsection{Effects of the low-frequency coherence neglecting the LFC}
\label{DeltaL=0}

We first analyze the SR of a nondegenerated $V$-system setting, as
a first step, the LFC to zero (see also the discussions in
references~\cite{Crubellier80,Crubellier85,Pavolini85}). In spite
of the fact that this assumption might be unphysical for an
ultrathin film (recall that $\Delta_L > \tau_R$), the analysis of
this ideal case will help us in understanding more complicated
situations with $\Delta_L \ne 0$.  Nondegeneracy means that the
magnitude of the splitting $\omega_{32}$ is larger than the
spectrum width of the SR in the presence of degeneracy. The latter
can be estimated on the basis of equations~(\ref{6Z})-(\ref{6tD})
as ${\tau_R^\prime}^{-1} = 4Z(0)\tau_R^{-1}$ for $\Delta_L = 0$.
Values of $\omega_{32}$ about several units of $\tau_R^{-1}$
suffice to model the outlined condition as $4Z(0) \le 2$.

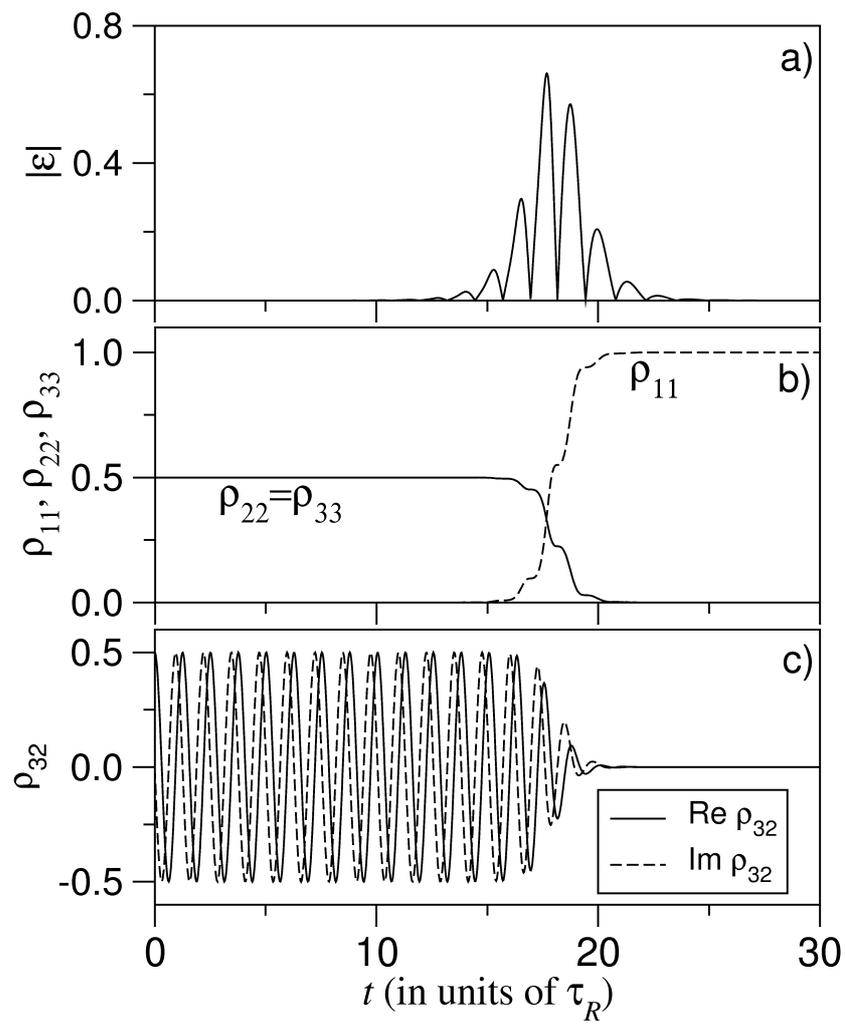
\begin{figure}[ht]
\caption{Kinetics of the SR field, $|\varepsilon| =
d|E|\tau_R/\hbar$, the level populations $\rho_{11}$, \;
$\rho_{22}$ and $\rho_{33}$, and the low-frequency coherence
$\rho_{32}$ calculated for zero LFC ($\Delta_L = 0$) at fixed
magnitude of the doublet splitting $\omega_{32} = 5\tau_R^{-1}$.
The initial conditions are: $\rho_{22}(0) = \rho_{33}(0) =
\rho_{32}(0) = 0.5$ and $R_{21}(0) = R_{31}(0) = 10^{-8}$.}
\label{fig2}
\end{figure}

Figure~\ref{fig2} shows the kinetics of the SR pulse and the level
populations calculated for $\omega_{32} = 5\tau_R^{-1}$ with the
following initial conditions: all the population is in the doublet
states, $\rho_{22}(0) = \rho_{33}(0) = 0.5$, and, additionally,
there exists a low-frequency coherence, $\rho_{32}(0) = 0.5$.
Within the subspace of states $|+ \rangle$ and $|- \rangle$, this
corresponds to the excitation of only the pure bright state $|+
\rangle$ ($\rho_{++}(0) = 1$), while all other populations and
coherences are equal to zero $\rho_{--}(0) = \rho_{11}(0) =
\rho_{+-}(0) = 0$ [see equations~(\ref{3R+1})-(\ref{3rho+-})]. As
is seen from figure~\ref{fig2}, the SR pulse deactivates {\it
completely\/} the state $|+ \rangle$. All the population is
finally transferred to the ground state $|1 \rangle$, as it takes
place in the case of two-level SR under the condition of total
inversion (see, for instance,
references~\cite{Gross82,Benedict96}). The modulation of the
kinetics with frequency $\omega_{32}$ is explained by the fact
that at nonzero splitting, the bright state $|+ \rangle$ is not a
stationary state: it periodically (with frequency $\omega_{32}$,
i.e.\@ rapidly in the scale of the SR) exchanges population with
the dark state $|- \rangle$. Indeed, keeping in
equations~(\ref{4rho++}),~(\ref{4rho+-}) and~(\ref{4rho--}) only
the terms with the density matrix elements within the subspace of
states $|+ \rangle$ and $|- \rangle$ and introducing the notations
$z = \rho_{++} - \rho_{--}$ and $y = i(\rho_{+-} - \rho_{-+})$, we
obtain
\begin{eqnarray}
        {\dot y} & = & - \omega_{32} \, z \ ,
\label{7y} \\
        {\dot z} & = & \omega_{32} \, y \ .
\label{7z}
\end{eqnarray}
These equations describe harmonic oscillations, with frequency
$\omega_{32}$, of a vector $(y,z)$ in the $YZ$ plane. Values $z =
1, -1$ correspond to the total population of bright and dark
states, respectively. The magnitude $y = 2\Im \rho_{+-}$ reflects
the low-frequency coherence. Note that $z^2 + y^2 =
\mathrm{const}$. For the initial conditions we are dealing with
($z (0) = 1$ and $y(0) = 0$), approximately one half of the period
of these oscillations the system  remains in the bright state
while the other half does not. This also explains why the delay
time of the SR in the present case ($t_D \approx 18\tau_R$) is two
times as large as compared to that time at $\omega_{32} = 0$.
Indeed, using the  above initial conditions in
equation~(\ref{6tD}), one obtains $t_D \approx 9\tau_R$ for a
degenerated doublet.

From the above discussion it is clear that changing the sign of
the initial low-frequency coherence,  i.e. setting $\rho_{32}(0) =
- 0.5$ or, in other words, $\rho_{--}(0) = 1$, will not  affect
the SR kinetics. After a half period of oscillations with
frequency $\omega_{32}$, the previous initial conditions will be
restored.

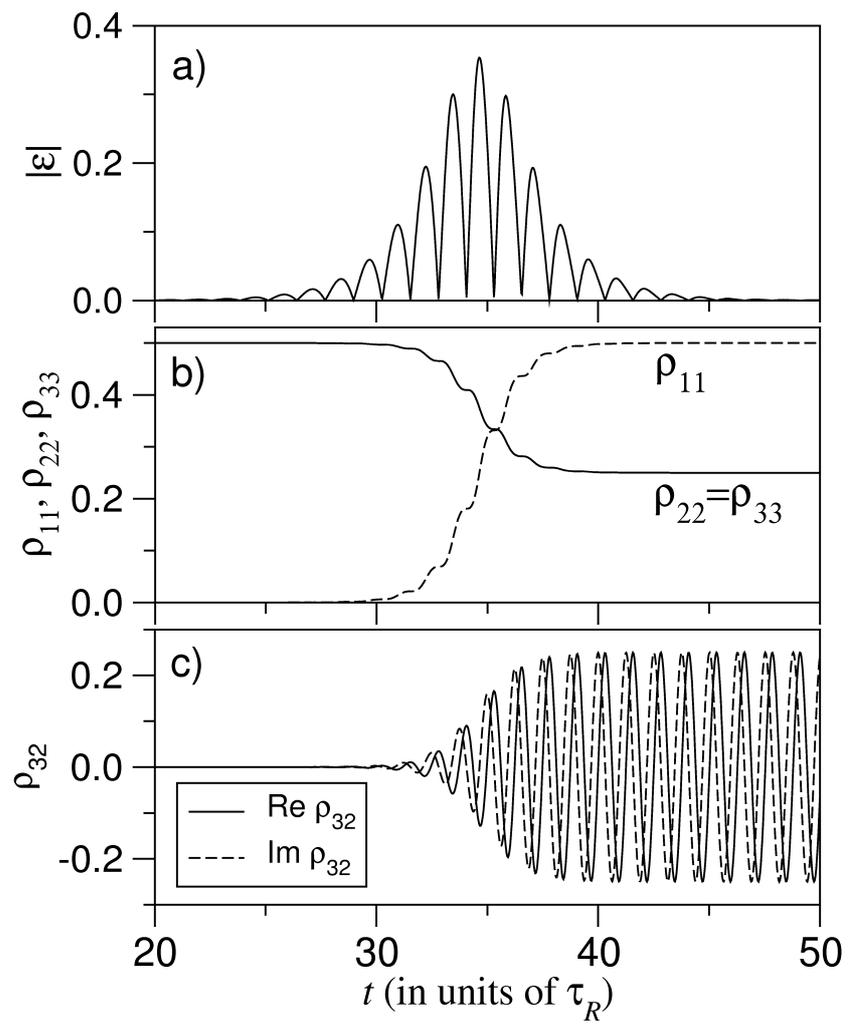
\begin{figure}[ht]
\caption{Same as in figure\protect{~\ref{fig2}}, except for
$\rho_{32}(0) = 0$.} \label{fig3}
\end{figure}

In figure~\ref{fig3} we depicted the SR kinetics calculated for
the same initial conditions as above, except for the low-frequency
coherence $\rho_{32}(0)$ was set to zero. Within the subspace of
states $|+ \rangle$ and $|- \rangle$, they now correspond to the
excitation of an incoherent mixture of the bright and dark states:
$\rho_{++}(0) = \rho_{--}(0) = 0.5$ and $\rho_{+-}(0) = 0$ [see
equations~(\ref{3rho++}),~(\ref{3rho--}) and~(\ref{3rho+-})]. One
can notice significant changes in the main features of the SR
pulse as compared to the previous case: both the delay time and
the pulse duration increased by approximately a factor of two and,
in addition, the doublet states remained equally populated after
the SR pulse has been emitted: $\rho_{22}(\infty) =
\rho_{33}(\infty) = 0.25$. More specifically, the population of
the bright state, i.e. only one half of the total population
accumulated in the upper states, is transferred to the ground
state during the SR. During the subsequent half remains trapped in
the dark state. The increase of the delay time and the duration of
the SR pulse by a factor of two is simply explained by the fact
that, in the present case, the initial value $Z(0) = \rho_{++}(0)
- \rho_{11}(0) = 0.5$ is twice as small as compared to the
previous situation. Recall that both $t_D$ and $\tau_R$ are
inversely proportional to $Z(0)$.

\subsection{Effects of the LFC ($\Delta_L \ne 0$)}
\label{DeltaL}

We turn now to studying the LFC effects on the SR kinetics.
Therefore, in what follows the magnitude of LFC, $\Delta_L$, is
regarded as a variable parameter while the doublet splitting,
$\omega_{32}$, is set to a fixed value. More specifically, we
present the results of numerical calculations for $\omega_{32} =
5\tau_R^{-1}$, meaning that the splitting is larger than the full
width of the SR spectrum which is $\approx 2\tau_R^{-1}$. For
small magnitudes of $\omega_{32}$ compared to $\tau_R^{-1}$, the
SR kinetics is well described by equations~(\ref{6Z})-(\ref{6tD}).
We also assume that the total initial population of the system,
$\rho_{22}(0) = \rho_{33}(0) = 0.5$ and $\rho_{11}(0) = 0$, with
equal populations of the doublet states, $\rho_{22}(0) =
\rho_{33}(0) = 0.5$. As we will show below, the output depends on
the initial value of the low-frequency coherence, $\rho_{32}(0)$,
as well. We restrict ourselves to two limiting cases:
$\rho_{32}(0) = 0$ and $\rho_{32}(0) = \pm \sqrt{\rho_{22}(0)
\rho_{33}(0)} = 0.5$. The former corresponds to the initial
excitation of incoherent mixture of the doublet states $|2
\rangle$ and $|3 \rangle$, while the latter implies the initial
excitation of the pure state $|+ \rangle$. The initial values of
the high-frequency coherences are set to $R_{21}(0) = R_{31}(0) =
10^{-8}$ as before.

\subsubsection{SR from an incoherent mixture ($\rho_{32}(0) = 0$)}
\label{rho32eq0}

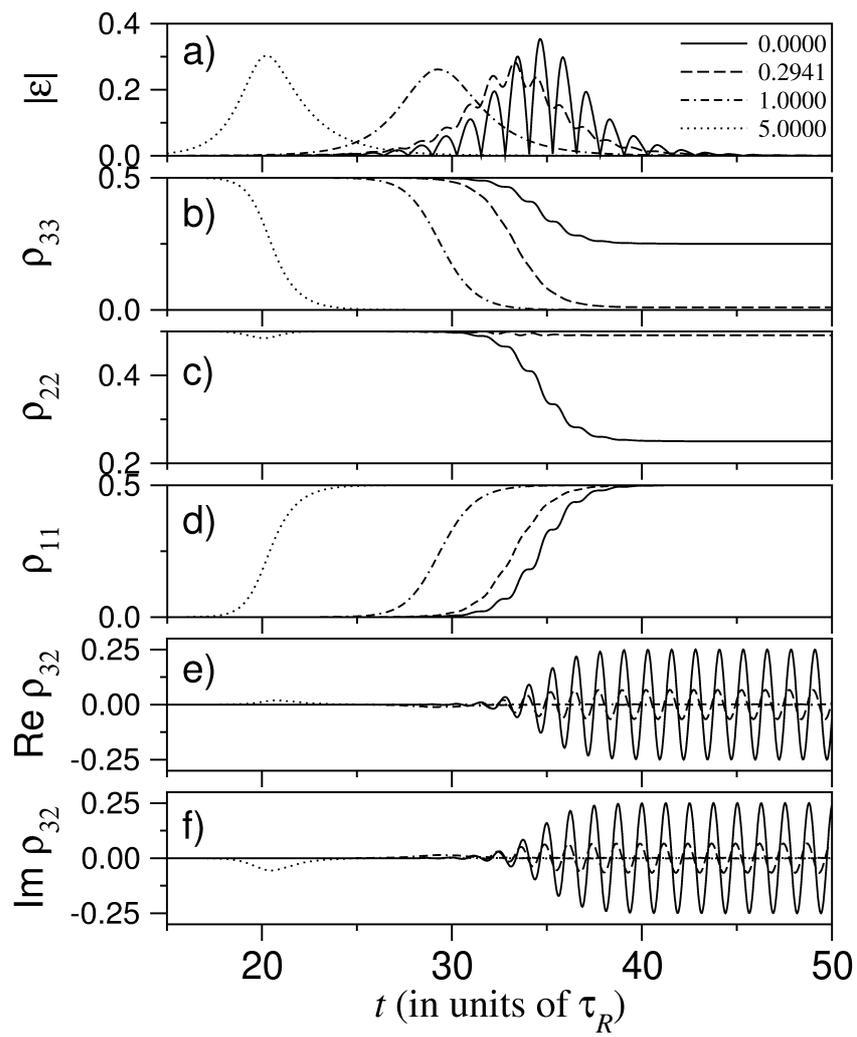
\begin{figure}[ht]
\caption{Effects of the LFC on kinetics of the SR field,
$|\varepsilon| = d|E|\tau_R/\hbar$, the level populations
$\rho_{11}$, \; $\rho_{22}$ and $\rho_{33}$, and the low-frequency
coherence $\rho_{32}$ calculated at a fixed doublet splitting
$\omega_{32} = 5\tau_R^{-1}$. The values of the LFC (in units of
$\tau_R^{-1}$ ) are given the panel $a)$. The initial conditions
are: $\rho_{22}(0) = \rho_{33}(0) = 0.5$ (total initial
inversion), $\rho_{32}(0) = 0$ (no initial low-frequency
coherence) and $R_{21}(0) = R_{31}(0) = 10^{-8}$.} \label{fig4}
\end{figure}

Figure~\ref{fig4} shows the SR kinetics calculated for
$\rho_{22}(0) = \rho_{33}(0) = 0.5$ (the total initial inversion)
in the absence of the initial low-frequency coherence,
$\rho_{32}(0) = 0$, varying the LFC magnitude, $\Delta_L$. As is
seen from this figure, increasing $\Delta_L$ affects drastically
the SR kinetics. While $\Delta_L$ is smaller than some ``critical"
value, the scenario of the SR is similar to that described in the
previous subsection, i.e., the transitions $| 2 \rangle
\rightarrow | 1 \rangle$ and $| 3 \rangle \rightarrow | 1 \rangle
$ evolve synchronously and the doublet states remain equally
populated after the SR pulse has been emitted: $\rho_{22}(\infty)
= \rho_{33}(\infty) = 0.25$. For larger $\Delta_L$, the transition
$| 2 \rangle \rightarrow | 1 \rangle$ in fact does not evolve,
conserving almost all the initial population in the state $|2
\rangle$, while the population of the state $|3 \rangle$ is
entirely transferred to the ground state $|1 \rangle$. This
explains the changes which occur in the SR kinetics on increasing
$\Delta_L$: disappearance of the oscillatory structure and
shortening both the SR pulse duration and delay time by
approximately a factor of two. Indeed, as $| 3 \rangle \rightarrow
 | 1 \rangle$ is the only transition contributing to the emission, the
problem is reduced to the two-level scheme with parameters
$\tau_R^\prime = 4Z(0)\tau_R = \tau_R$ and $t_D = - \tau_R^\prime
\ln R_{31}(0) \approx 17\tau_R$, which are characteristic for the
two-level SR. It worth mentioning that the suppression of the
transition $| 2 \rangle \rightarrow | 1 \rangle$ occurs at $d_{21}
= d_{31}$, i.e., under the condition of equivalent coupling of the
individual transitions to the field.

The physics of such a behavior is as follows. A $V$-atom
represents two transitions coupled to each other by the common
field which includes the emission term ($\sim \tau_R^{-1}$) and
the LFC ($\sim \Delta_L$) [see equation~(\ref{1Eprime})]. In the
presence of population inversion, the amplitudes of optical
oscillations, $R_{21}$ and $R_{31}$, grow in time. Their
increments are equal to each other, in the absence of the LFC and
under the condition $\mu_{13} = \mu_{12}$ and $\rho_{33}(0) =
\rho_{22}(0) = 0.5$. However, they become different after
appearing the LFC (see below). This makes the transition $| 3
\rangle \rightarrow  | 1 \rangle$ to evolve faster than $| 2
\rangle \rightarrow | 1 \rangle$. The final state of the system
depends on the relationship between $R_{31}$ and $R_{21}$ at those
times when the SR pulse is already well developed, i.e., at $t
\approx t_D$. If $|R_{21}(t_D)| \approx |R_{31}(t_D)|$ then both
transitions still evolve synchronously, while at $|R_{21}(t_D)|
\ll |R_{31}(t_D)|$ the initial population of level $| 3 \rangle$
($0.5$ in our case) is transferred to the level $| 1 \rangle$
before the oscillations $| 2 \rangle \rightarrow | 1 \rangle$
begin to built up. It makes the population inversion between
levels $| 2 \rangle$ and $| 1 \rangle$ equal to zero and thus
prevents the superradiant evolution of this channel which explains
the disappearance of the oscillations of the SR pulse.

An analysis of the linear stage of the emission, i.e., keeping all
the quantities equal to their initial values, except for $R_{21}$
and $R_{31}$, provides a solid support to the above arguments.
Equations~(\ref{2R31})-(\ref{2rho11}), linearized with respect to
$R_{21}$ and $R_{31}$ and adapted to the conditions used in the
numerical simulations ($\mu_{31} = \mu_{21} = 1$, \ $\rho_{33}(0)
= \rho_{22}(0)$, \ $\rho_{11}(0) = \rho_{32}(0) = 0$), have the
form
\begin{eqnarray}
{\dot R}_{31}  & = & \left[ -i\frac{\omega_{32}}{2} + \left(
\frac{1}{\tau_{R}}
        - i\Delta_{L} \right)W \right] R_{31}
        + \left( \frac{1}{\tau_{R}} - i\Delta_{L} \right)W R_{21} \ ,
\label{8R31} \\
{\dot R}_{21} & = & \left[\> \> \> i\frac{\omega_{32}}{2}+\left(
\frac{1}{\tau_{R}}
        - i\Delta_{L} \right)W \right] R_{21}
        + \left( \frac{1}{\tau_{R}} - i\Delta_{L} \right)W R_{31} \
        ,
\label{8R21}
\end{eqnarray}
\\
where $W \equiv \rho_{33}(0) - \rho_{11}(0) = \rho_{22}(0) -
\rho_{11}(0)$. Solving these coupled equations is straightforward.
Below, we write down the solution in the limit $\tau_R^{-1},
\Delta_L \ll \omega_{32}$:
\begin{equation}
        R_{21} \simeq R_0 e^{\lambda_1 t} \ , \qquad
    R_{31} \simeq R_0 e^{\lambda_2 t} \ ,
\label{R21R31}
\end{equation}
where
\begin{equation}
        \lambda_{1,2} = i \left(\pm \frac{\omega_{32}}{2} - \Delta_L W \right)
        + \frac{W}{\tau_R}\left(1\mp 2W \frac{\Delta_L}{\omega_{32}} \right) \
        .
\label{1lambda12}
\end{equation}
As seen from equation~(\ref{1lambda12}), the increment of $R_{31}$
is indeed larger than that for $R_{21}$, i.e. $R_{31}$ grows
faster than $R_{21}$. Thus, during the linear stage of the SR
\begin{equation}
        \frac{|R_{31}|}{|R_{21}|} =
        \exp\left( 4W^2 \frac{\Delta_L}{\omega_{32}}\frac{t}{\tau_R} \right) \ .
\label{Ratio1}
\end{equation}
Recall that the linear solutions for $R_{31}$ and $R_{21}$ are
valid almost up to the SR pulse maximum (see, for instance,
reference~\cite{Malyshev00}). Then, applying this formula for $t =
t_D$ and equating the exponent to unity, one obtains an estimate
for $\Delta_L^c$,
\begin{equation}
        \Delta_L^{c} = \frac{\omega_{32}}{4W^2} \frac{\tau_R}{t_D} \ ,
\label{1DeltaLc}
\end{equation}
which separate two regimes of the SR. At $\Delta_L <
\Delta_L^{c}$, both transitions evolve synchronously, while for
the opposite sign of the inequality, the transition $| 2 \rangle
\rightarrow  | 1 \rangle$ is blocked for the reasons discussed
above.

Concerning numerical data ($\omega_{32} = 5\tau_R^{-1}$,  $W =
0.5$,  and $t_D \approx 35\tau_R$), equation~(\ref{1DeltaLc})
yields $\Delta_{L}^{c} = \omega_{32}/35 = (1/7) \tau_R^{-1}$. This
estimate is in good agreement with the numerical data (see
figure~\ref{fig4}). We stress that the estimated value of
$\Delta_L^{c}$ is smaller than the half-width of the SR spectrum
given by $\tau_R^{-1}$. In other words, even if the LFC (the
dynamical resonance frequency shift) are spectroscopically hidden
due to the natural pulse broadening, it drastically affects the SR
kinetics.

It is to be noticed that the above behavior of the SR of a
$V$-system resembles the peculiarity of the SR for the
$\Lambda$-arrangement of nondegenerated
levels~\cite{Zaitsev99b,Zaitsev01,Ryzhov01,Malyshev00}, namely in
the presence of the LFC, all the population from the upper level
of the $\Lambda$-system is transferred to the lower level of the
doublet, while the higher doublet level remains unpopulated after
the SR pulse has gone.

\subsubsection{SR from the pure state ($\rho_{32}(0) = 0.5$)}
\label{rho32ne0}

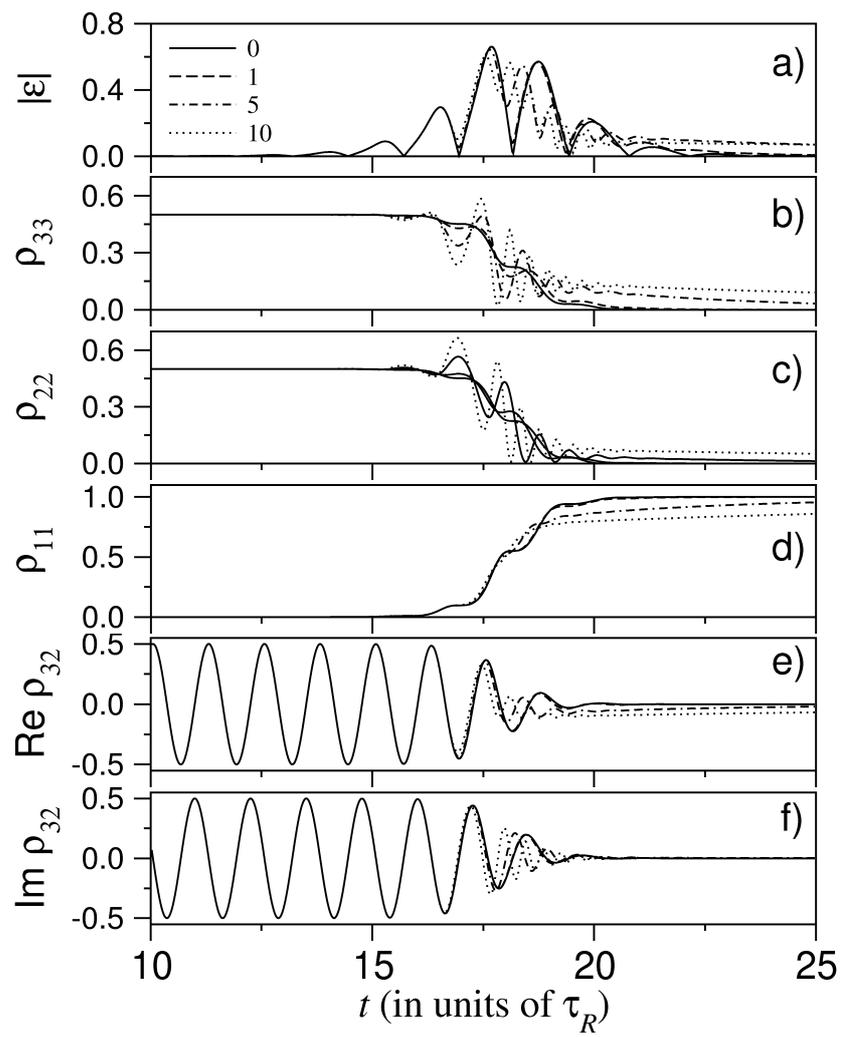
\begin{figure}[ht]
\caption{Same as in figure\protect{~\ref{fig4}}, except for
$\rho_{32}(0) = 0.5$.} \label{fig5}
\end{figure}

Figure~\ref{fig5} shows  the effects of the LFC on the SR kinetics
obtained for $\rho_{22}(0) = \rho_{33}(0) = \rho_{32}(0) = 0.5$
or, in other words, when initially the pure bright state
$|+\rangle$ is fully populated, $\rho_{++}(0) = 1$, while
$\rho_{11}(0) = \rho_{--}(0) = \rho_{+-} = 0$. As is seen from
figure~\ref{fig5}a, the present case differs noticeably from the
one where initially no low-frequency coherence is created
($\rho_{32}(0) = 0$, see previous section). First of all, the LFC
does not affect the SR delay time at all. Therefore, the linear
stage of the SR is unuseful here in predicting the changes in the
SR kinetics, as they occur when the nonlinearity is already well
developed.

The changes concern the second half and final stage of the pulse.
The pulse shows an oscillatory structure which now cannot be
associated with $\omega_{32}$-oscillations. The frequency of the
oscillations grows upon increasing $\Delta_L$ and, in fact,
reflects the magnitude of the latter. The populations of the
doublet states also undergo antiphased (with respect to each
other) oscillations at the same frequency as the pulse does. This
indicates that the doublet states start to exchange the population
when the SR pulse is developed. A qualitative interpretation of
this effect is as follows. Recall that the LFC shifts the
frequency of the transitions $| 2 \rangle \rightarrow | 1 \rangle$
and $| 3 \rangle \rightarrow | 1 \rangle$ by $\Delta_L (\rho_{22}
- \rho_{11})$ and $\Delta_L (\rho_{33} - \rho_{11})$,
respectively. Initially, these shifts are equal to each other.
Figure~\ref{fig5}b shows that the transition $| 3 \rangle
\rightarrow | 1 \rangle$ start to develop first. This reduces the
initial detuning, $\omega_{32}$, between the transitions. As a
result, the radiation, which is emitted via the transition $| 3
\rangle \rightarrow | 1 \rangle$, is absorbed by the transition $|
2 \rangle \rightarrow | 1 \rangle$. It further reduces the
detuning, stronger for larger $\Delta_L$. After that, the
transition $| 2 \rangle \rightarrow | 1 \rangle$ starts to emit
while $| 3 \rangle \rightarrow | 1 \rangle$ to absorb, i.e.\, the
transitions exchange their role.

The initial population is mostly transferred to the ground state,
as it takes place for the same initial conditions at $\Delta_L =
0$. However, a small part of the population remains trapped in the
dark state $|- \rangle$, unlike the case of $\Delta_L = 0$. This
is the reason why the SR pulse has a long tail.

\section{Summary and concluding remarks}
\label{Summary}

We studied theoretically the SR from an ultrathin film of $V$-type
atoms taking into account the LFC to the average Maxwell field. We
show that the interplay between the doublet splitting,
low-frequency coherence (within the subspace of the doublet
states) and LFC may significantly affect the scenario of SR.
Several conclusion can be drawn from our results:

(i) Under the condition of degeneration, the three-level problem
is equivalent to that for a two level system with a renormalized
SR time. The role of the LFC is also similar to that for the
two-level problem and manifests itself as a phase modulation of
the SR pulse.

(ii) For a nondegenerated $V$-system, the LFC correction affects
drastically the SR scenario, allowing to develop one of the
transitions and blocking the other one under specific conditions.
This effect may occur even if the LFC is small compared to the SR
spectrum, i.e., when the LFC is spectroscopically hidden.

(iii) The SR scenario is sensitive to the amount of low-frequency
coherence (within the subspace of the doublet states) as well as
to the magnitude of the total inversion, thus providing a way to
control the SR regimes.

To conclude, we discuss the conditions required to prove
experimentally the predicted regimes of the $V$-type SR. First of
all, one should look for a dense ensemble of radiators where the
LFC is large compared to the line width. In
reference~\cite{Crenshaw92a}, O$_2^{-}$ ions in KCl:O$_2^{-}$
crystals and bound I$_2$ excitons at donor sites in CdS single
crystals were considered suitable for observing the LFC effects.
In relation to our model, one should bear in mind that in
disordered ensembles of dipole radiators, like the case of
O$_2^{-}$ centers and bound I$_2$ excitons, the LFC fluctuates.
The average LFC drives the level shifts, while the fluctuating
part contributes to the dipole-dipole line broadening. It turns
out that both shift and line width are of the same order of
magnitude~\cite{Manassah83}, as it takes place in dense gas
systems~\cite{Friedberg89}. Because of this fact, these systems
can hardly present the effects we are discussing.

Thin films of some organic compounds, such as naphtalene and
antracene, as well as materials containing unoccupied $d$ or $f$
orbitals, such as Cr$_2$O$_3$ or MnO$_2$, might be promising
materials for this task.  As they are crystalline, the
intermolecular dipole-dipole interaction (and, subsequently, the
LFC) does not fluctuate. Furthermore, at low temperatures, the
optical excitations in these materials are Frenkel
excitons~\cite{Davydov71,Agranovich82,Rashba82}. This fact implies
that the intermolecular dipole-dipole interaction (the LFC, in
other words) dominates over dephasing. As the density of the
optically active units in crystals is generally high ($N_0\sim
10^{21}-10^{22}\,$cm$^{-3}$), the SR time constant $\tau_R$ may be
small compared to the dephasing time. Indeed, $\tau_{R} =
\hbar\lambda_c/(2\pi)^2 d^{2}N_{0}L =
(8\pi/3)(N_0\lambda_c^3)^{-1} (\lambda_c/L)\tau_0$, where
$\lambda_c = 2\pi/k_c$ and $\tau_0 = 3\hbar/4d^2k_c^3$ is the
spontaneous emission time of a single emitter. Let us take
$\lambda_c = 5 \times 10^{-5}\,$cm and assume that the transitions
are dipole allowed ($\tau_0 \sim 10^{-8}s$), which is typically
the case for organic crystals. Then, for a film thickness $L =
0.1\lambda_c$ we estimate $\tau_R$ as being of the order of
$10\,$fs. The exciton absorption line width is typically about few
hundreds cm$^{-1}$. This gives $1\,$ps as an estimate for the
dephasing time, that is, hundreed times longer than $\tau_R$. On
the other hand, vibronic structure of aromatic crystals seems
suitable for forming a $V$-configured system. The SR of high
density Frenkel excitons was observed in single organic crystals
of R-phycoerythrin molecules at room temperature \cite{Wang95}.
The SR pulse was found to be phase-modulated,  thus indicating the
relevance of the LFC. Therefore, R-phycoerythrin single crystals
are promising candidates to prove the effects predicted in this
work.

\ack

V. A. M. acknowledges the financial support through a NATO
Fellowship, and la Universidad Complutense de Madrid for
hospitality. F. D-A. was supported by DGI-MCyT (Project
MAT2000-0734) and CAM (Project 07N/0075/2001). F. Carre\~{n}o, M.
Ant\'on and O.G. Calder\'on were supported by project no
BFM2000-0796 (Spain).

\Figures

\end{document}